# Analysis of Polar Motion Variations from 170-year Observation Series

## N. Miller, Z. Malkin

Pulkovo Observatory, St. Petersburg, Russia

**Abstract.** This work is devoted to investigation of low frequency variations in Polar motion (PM). It has been shown that the main PM features can be effectively investigated using not only time series of the Pole coordinates, but also using series of latitude variations obtained from observations at one observatory. Such an approach allows us to increase the length of observation series available for analysis. In our study, we extended the IERS PM series back to 1840. We investigate trends and (quasi) harmonic oscillations with periods from one year to decades. The main results were obtained making use of the Singular Spectrum Analysis. Other methods were also used for specific analysis and independent check. The most interesting results are detection of two new large phase jumps in the 1840s and 2000s, and revealing of 80-year period in the PM variations.

**Introduction**

The Chandler wobble (CW) is one of the main components of motion of the Earth's rotation axis relative to the Earth's surface, also called Polar motion (PM). The CW is one of the main eigenmodes of the Earth rotation, and investigation of its properties such as period, amplitude and phase variations is very important for the understanding of the physical processes in the Earth, including its surface, interior, atmosphere and ocean. Since the discovery of CW, numerous papers were devoted to analysis of this phenomenon.

The "main" phase jump in the 1920s remains to be the most interesting feature of the CW, which, in particular, is a real test of Earth rotation theories. It should be noted that most of papers devoted to the CW studies are based chiefly on the observations of the Earth rotation obtained in the period from 1899 (start of the International Latitude Service) till the end of the 1990s. In this study we used the whole time span of the IERS (International Earth Rotation and Reference Systems Service) PM time series C01 from 1846 till 2010. Our previous investigations of this series have shown clear evidence of two other large CW phase jumps that occurred in the beginning and the end of the interval covered by the C01 series [1, 2]. In this paper we used a little bit longer time series, which allowed us to better analyze the CW behavior in the 2000s.

As to CW variations in 1840s, an overview of literature is given in [3]. Many authors analyzed CW parameters variations for the period 1840–1860, e.g., S. Chandler, A. Ivanov, B. Wanach, H. Kimura, S. Kosnisky, A. Vasiliev, A. Orlov, N. Sikeguchi. Their results showed decreasing of the CW amplitude down to about 008" and a change in the CW phase. In this paper we investigated a possibility to use latitude variations of Pulkovo Observatory since 1840 to improve CW analysis at this interval.

To improve reliability of the results, several methods of analysis were used, and their results corroborate the common conclusion on the two existing large CW phase jumps in the 1850s and nowadays preliminary found in our previous works.

**Data used**

The IERS C01 series[1] is the longest PM series publicly available starting from 1846. It is computed by combination of observations made at many observatories using various observational techniques from classical latitude observations to modern space geodesy results, and processed using various methods. For this reason, this series is not homogeneous with respect to both systematic and random errors. This series is given at 0.1 year interval for XIX century and 0.05 year interval at XX century. Because of the method used for computation of the C01 series, it is published with delay up several months. So, the IERS C04 series updated twice a week was used to prolong the IERS PM data up to the date.

On the other hand, as was shown in [4], a series of latitude variations at a single observatory can be effectively used for analysis of the CW amplitude and phase variations. At Pulkovo, two instruments, Repsold's transit instrument installed in the prime vertical (TIPV) and Ertel's vertical circle (VC), started ob-servations in 1840. The main goal of these observations was determination of absolute declination of stars. However, during these observations the latitude was be also determined and can be used for PM studies. These observations allowed us to prolong the Pulkovo latitude series back to 1840 [4]. Figures 1 and 2 show the latitude variations obtained with these two instruments, and Fig. 3 shows the combined series.

---

[1] http://hpiers.obspm.fr/eop-pc/index.php?index=C01&lang=en

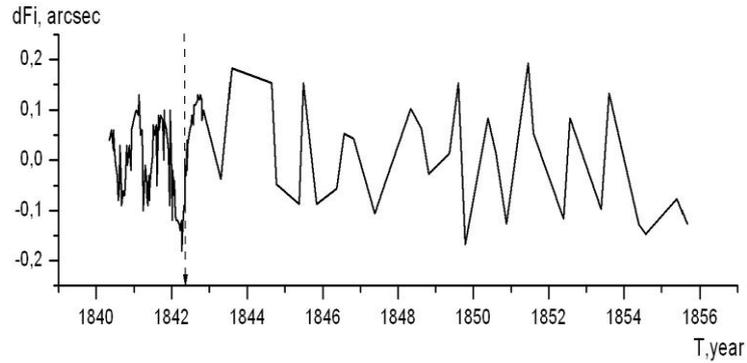

Fig. 1. The latitude observations with the transit instrument in the prime vertical.

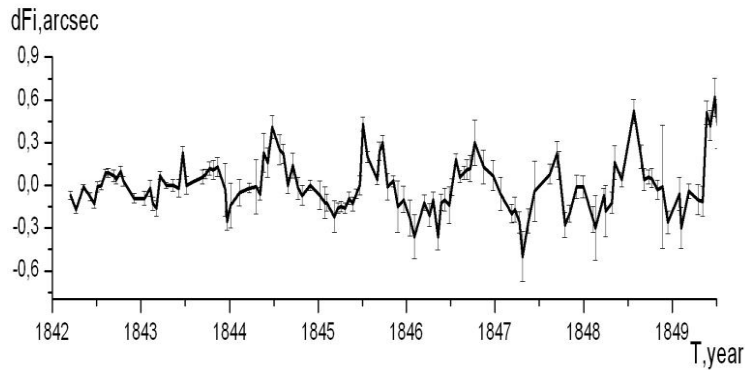

Fig. 2. The latitude observations with the vertical cycle.

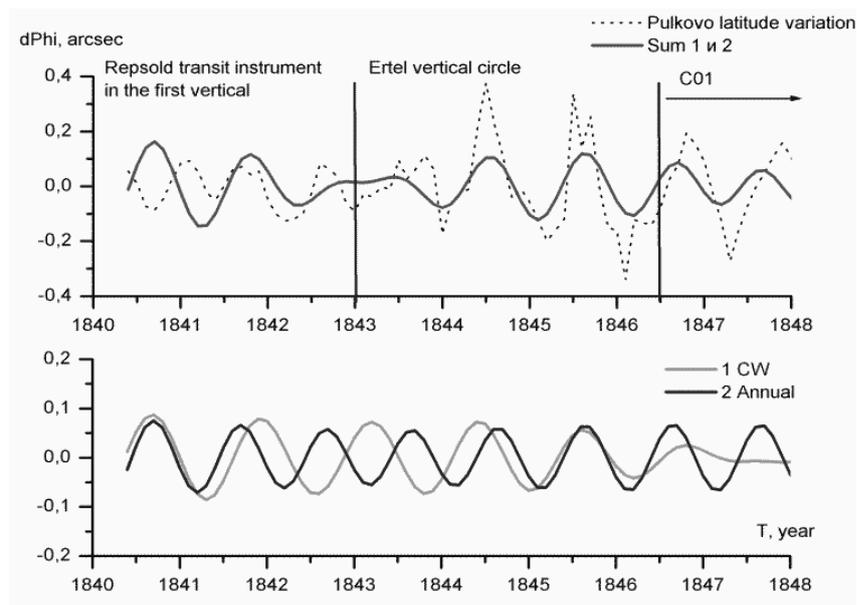

Fig. 3. Pulkovo latitude variations for 1840-1848 used to prolong the latitude variations back (top) and corresponding CW and annual components.

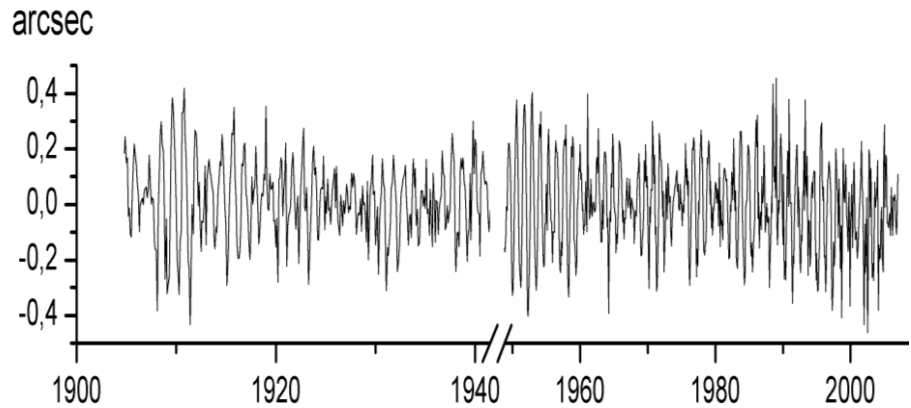

Fig. 4. The Pulkovo latitude variations obtained with ZTF-135.

The combined series consist of PITV series for 1840.3–1843.0, VC series for 1842.4–1846.5, and C01 series for 1846.0-1848.0. As was shown in [3], these series agree well at the interval of intersection 1842.4–1843.0. So, they were simply averaged at the overlapping intervals 1842.4-1843.0 для PITV и VC series and 1846.0-1846.5 для VC и C01 series. The lower part of Fig. 3 shows the annual and the CW signal extracted from combined series (see below description of methods used for this purpose).

In 1904, the new large zenith-telescope ZTF-135 was installed at the Pulkovo Observatory. The telescope was constructed in Pulkovo by G. A. Freiberg-Kondratiev and during his long history proved to be one of the most accurate instrument for latitude observations. ZTF-135 provided a unique series of the latitude of the Pulkovo observatory from 1904 till the end of 2006 with 7-year interruption during the 2nd World War in 1941–1948. During this period, 36 observers obtained about 170,000 latitude observations. The Pulkovo latitude variations derived from these observations are presented in Fig. 4

Finally, using all the available data, we constructed a combined Pulkovo latitude time series from 1840.3 till 2011.4 with 0.1 year step.

**Methods of analysis**

Our goal was to decompose the pole coordinates time series C01+C04 and Pulkovo latitude time series into principal components such as trend, CW and annual term. Other components found in the observed series are of much lesser amplitude and are not considered here. The main method of our analysis was singular spectral analysis (SSA). Several other methods were used for auxiliary tasks and for testing of main results obtained with SSA.

SSA belongs to the class of methods of natural orthogonal functions, for which basis functions are determined by the data themselves. The SSA method provides good resolution in both time and frequency domain and allows us to analyze time series with complex structure. It allows also to obtain the time variations of the amplitude, frequency and phase of the signal under investigation. Besides this method s capable to reliable discriminate a signal and random (high-frequency) noise, even if the noise level is changed with time. The latter is especially useful for analysis of observational data. As shown in previous studies, it can be effectively used in investigations of the Earth rotation [4, 5].

Digital band-pass filtering can also be effectively used for extracting of CW signal [2]. In this work, the elliptic filter of 5th order was used as realized by the Ellip function of the Matlab Signal Processing Toolbox). The obtained results are very similar to ones obtained with other filters used in [2]. Other methods used in this study for CW analysis were Fourier transform and wavelet analysis (with Morlet basis).

The amplitude and phase variations were investigated using the complex Hilbert transformation as realized by the Hilbert function of the Matlab Signal Processing Toolbox.

**Principal PM components**

Using the SSA analysis, we extracted principal components of the Pulkovo latitude variations: non-linear trend, annual and CW signals as shown in Fig. 5. The CW signal was also extracted from this series using the elliptic digital filter with the window 1.19±0.1 yr, and from the C01 time series using both SSA and elliptic digital filter techniques (Fig. 6). Figure 7 presents the power spectra of the time series exposed in Fig. 6.

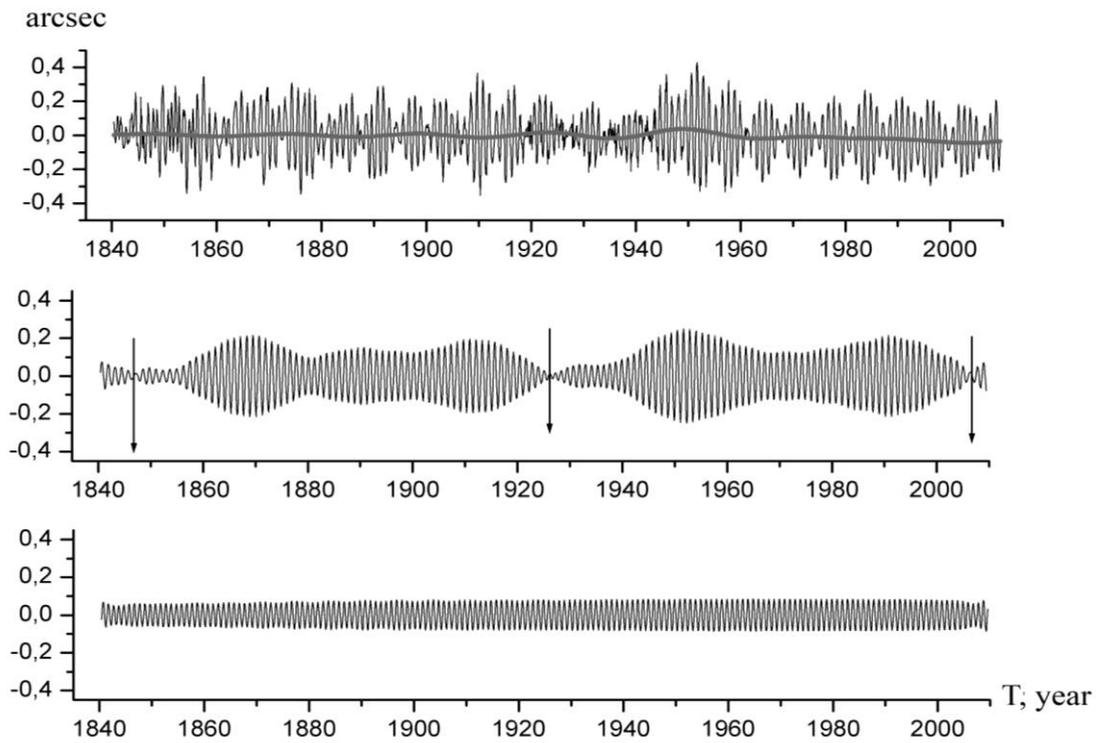

Fig. 5. Combined Pulkovo latitude series for 1840.4-2011.4 with non-linear trend and extracted CW and annual components (*from top to bottom*).

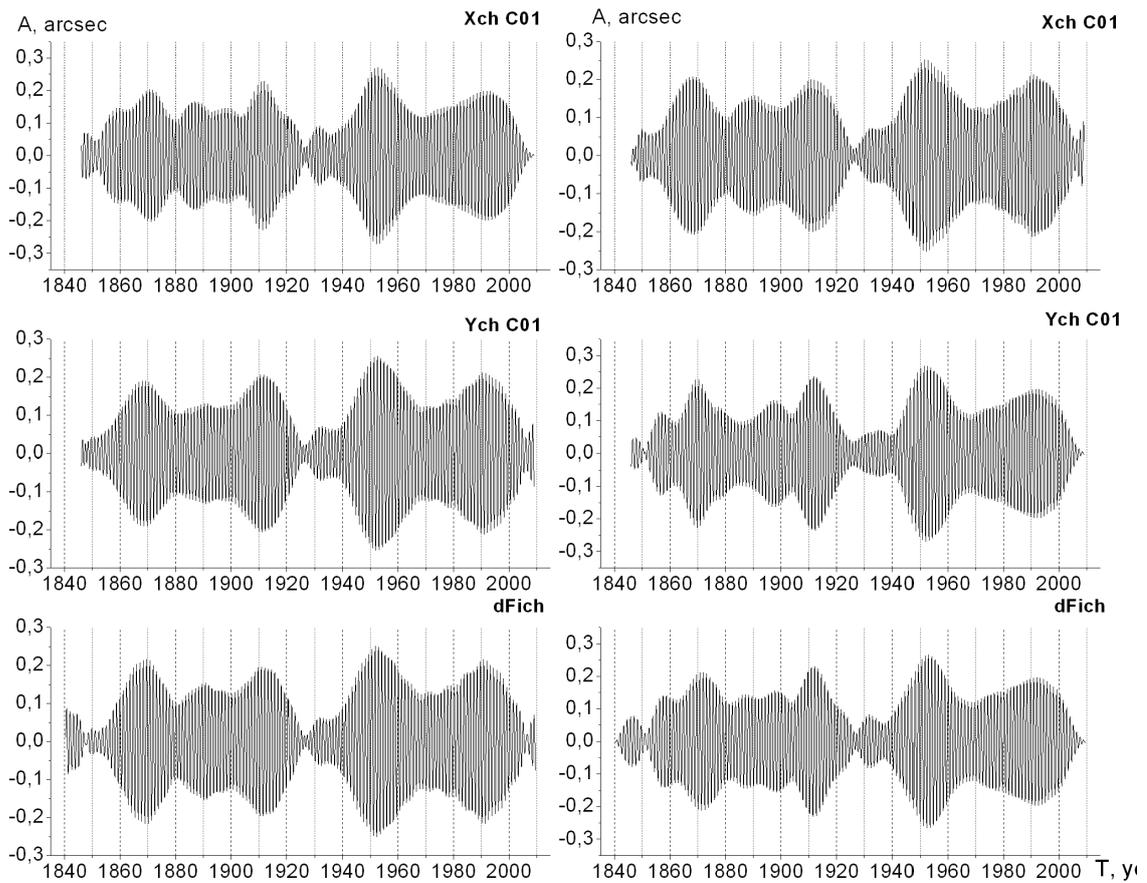

Fig. 6. The CW components in the C01 Pole motion (Xch and Ych) and Pulkovo latitude variations (dFich) obtained by the SSA method (*left*) and by elliptical digital filtering (*right*).

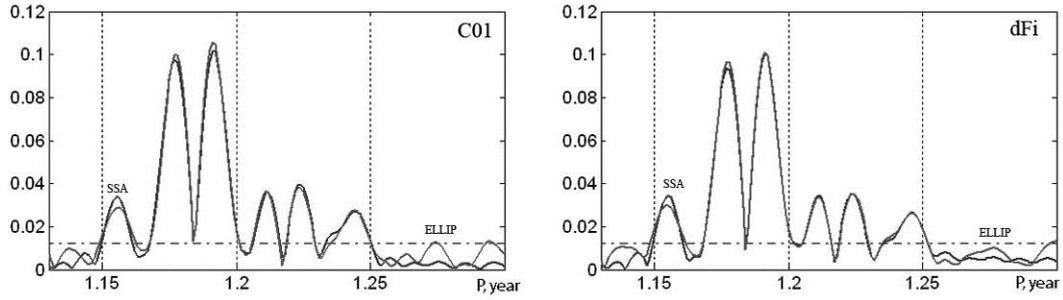

Fig. 7. Power spectra of the time series presented in Fig. 6. The horizontal line shows the 95% confidence level of the spectrum amplitude. Detailed discussion of several spectral peaks is given in [4]. Amplitudes of harmonics are given in arcseconds.

One can see from Figs. 6 and 7 that the CW signal looks quite similar in all filtered series with some small differences near the ends of the interval. The spectra of both series are also practically identical.

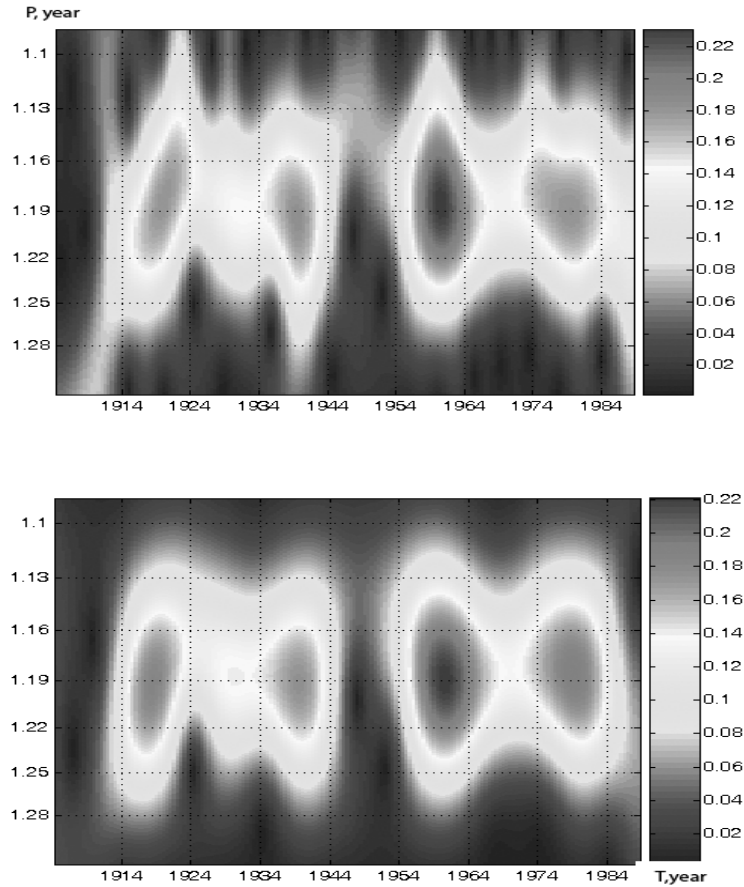

Fig. 8. The wavelet transforms of the original Pulkovo latitude variations time series (*top*) and the CW component extracted with the help of SSA (*bottom*).

The wavelet transform (with the Morlet basis) was used for analysis of the initial series of the Pulkovo latitude variations and its CW component derived with the help of SSA (Fig. 8). It is seen that the CW component extracted from the initial time series have the structure more pronounced than the CW in the initial series.

Figure 9 presents the CW signal divide into two intervals of similar behavior, 1846.8–1926.2 and 1926.2–2005.6, which can be observed in Fig. 6. Figure 10 presents spectra of CW signal computed for both intervals. From these plots one can clearly see that the CW amplitude variations are very similar for both intervals. This result can provide an evidence of a new CW period of 79.4±0.2 yr.

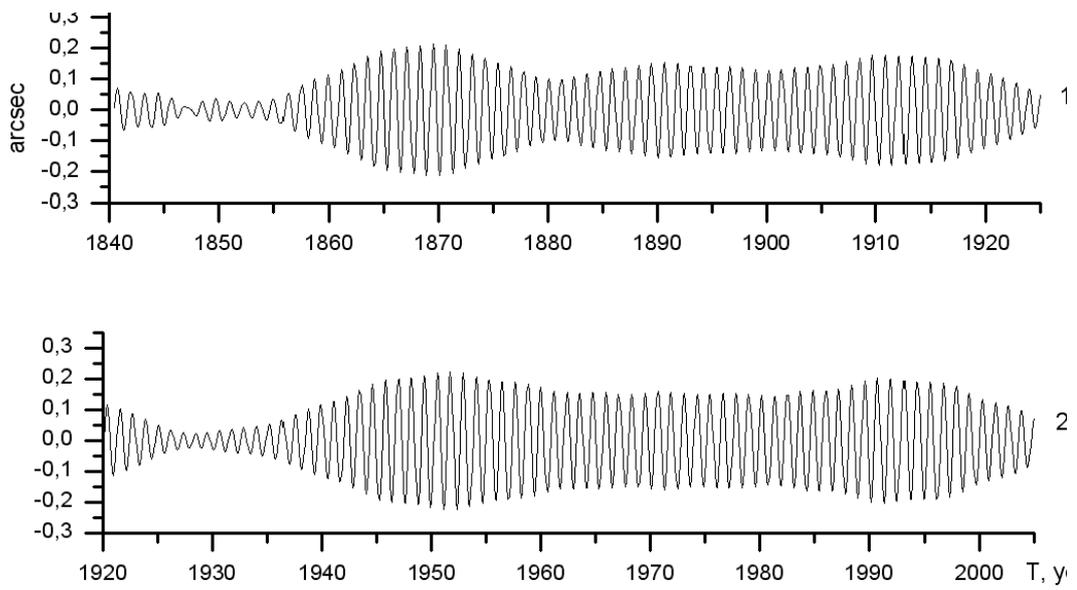

Fig. 9. Two similar intervals of CW variations.

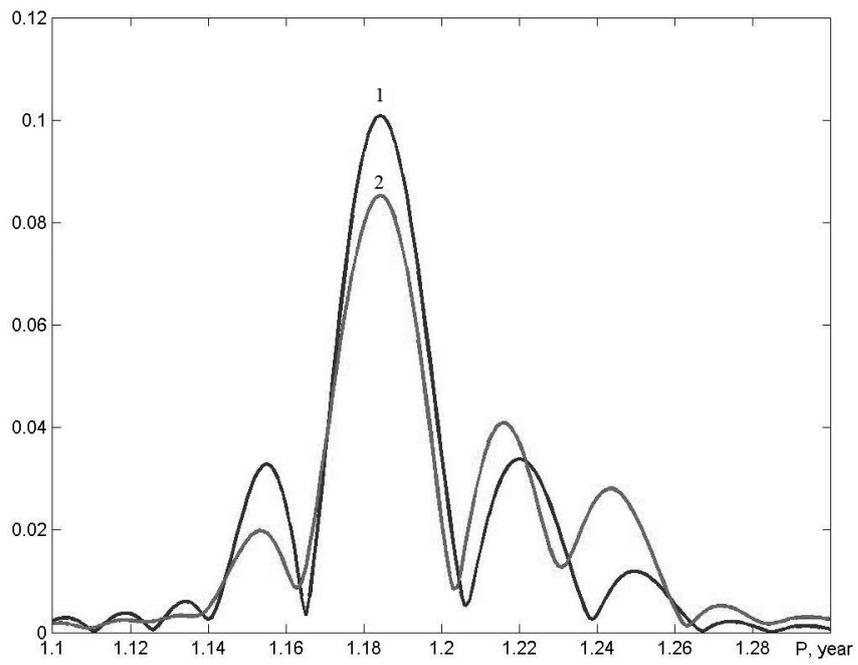

Fig. 10. The power spectra of the two similar intervals of CW variations.

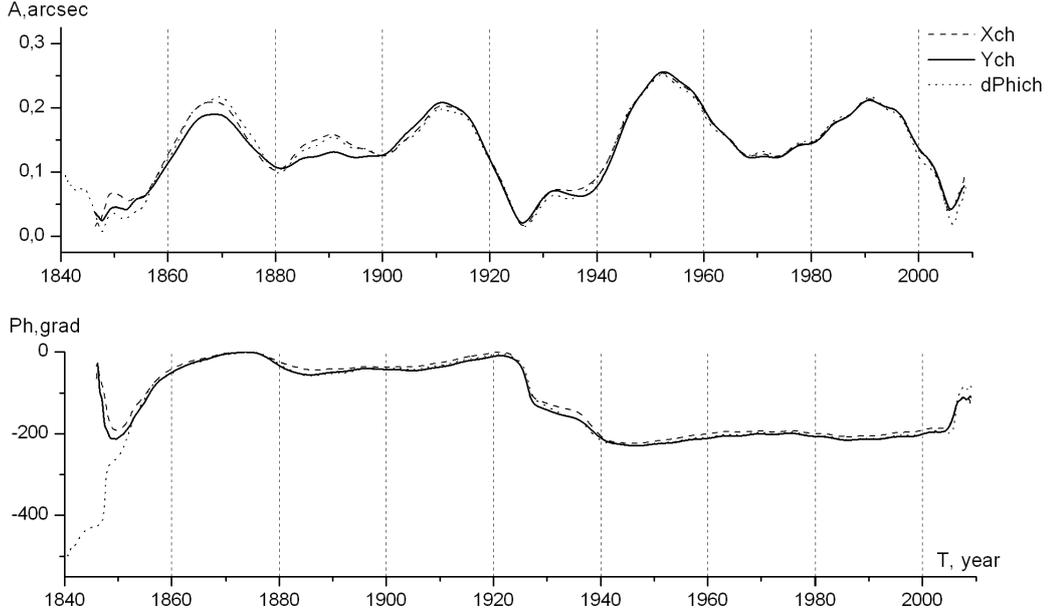

Fig. 11. The CW amplitude and phase variations computed with the SSA.

Detailed investigations of the CW amplitude and phase variations were performed using the following method. Let us consider a general CW model with the time-dependent amplitude and phase

$$Xp(t) = A(t)\cos\varphi(t), \ Yp(t) = A(t)\cos\varphi(t), \qquad (1)$$

where Xp and Yp are the Pole coordinates. The CW amplitude time series can be easily computed as

$$A(t) = \sqrt{Xp(t)^2 + Yp(t)^2} \ . \qquad (2)$$

In contrast to the computation of the CW amplitude, the calculation of the CW phase is not such an unambiguous procedure. Hilbert transform methods were applied to evaluate the CW phase variations. Thus obtained the CW amplitude variations are shown in Fig. 11 for two CW series. We can see that both CW series show very similar behavior of the CW amplitude, with some differences near the ends of the interval. In both CW series, three deep minima of the amplitude below 0.05 mas around 1850, 1925 and 2005 are unambiguously detected.

**Summary and conclusions**

In this paper, we have investigated the 165-year IERS PM series available from IERS and 171-year series of the Pulkovo latitude variations obtained from combination of different data. To improve the reliability of the results we used several analysis methods. To extract the CW signal from this series, SSA-based and digital filters were applied. Thus obtained two CW series were used to investigate the CW amplitude and phase variations.

All the methods used gave very similar results, with some differences at the ends of the interval. These discrepancies can be explained by different edge effects of the methods used, and have shown existence of two epochs of deep CW amplitude decrease around 1850 and 2005, which are also accompanied by a large phase jump, like well-known event in 1920s.

Unfortunately, both periods of the phase disturbances found in this paper are located at the edges of the interval covered by the observations. As to the end of the interval, the next decade will allow us to quantify the phase jump in the beginning of the 21th century.

On the other hand, supplement study seems to be extremely important to improve our knowledge about PM in XIX century, including an extension of the PM series in the past. In particular, as investigated in detail by Sekiguchi [6], there are several latitude series obtained in the first half of 19th century, which can be used to extend the IERS C01 PM series back to 1830s. However, most of the observations are of rather poor quality to be used directly in computation of an extended PM series. Clearly, this material worth revisiting and reprocessing using HIPPARCOS and maybe later GAIA star positions

Two repeated similar CW structures in 1846.8-1926.2 and 1926.2-2005.6 were found; and a new period of about 80 years in the CW variations can be suspected.